\documentstyle[aps,prl,multicol,epsf]{revtex}
\input epsf
\begin{document}
\draft
\widetext
\title{Superconductors of mixed order 
parameter symmetry in a Zeeman magnetic field}
\author{Haranath Ghosh} 
\address{Department of Physics, University of Arizona, Tucson, AZ 85721, USA.}
\date{\today}
\maketitle
\begin{abstract}
We study the effect of Zeeman magnetic field in the
superconducting phase with two component order parameter 
scenario, such as, $\rm d_{x^2-y^2} + e^{i\theta}{\alpha}$, where
$\alpha = d_{xy},~ s$. 
This scenario is 
equivalent to applying magnetic field parrllel to $\rm Cu-O$ planes.
In a weak magnetic field, which does not cause
much change to the predominant $d$-wave, supresses the minor 
$\alpha$ component leading to pure $d$-wave phase. 
This observation is in contrary to the effect of magnetic field applied in
the $c$-direction to the $\rm Cu-O$ planes which is believed to
induce a minor component $\alpha$ to $d$-wave superconductors.
  We also show that the response of such superconductors
to a weak Zeeman magnetic field can be quite different depending 
on the phase $\theta$ of the minor component ($\alpha$).
\end{abstract}
\pacs{74.25.Nf,74.25.Dw,74.62.-c}
\begin{multicols}{2}

\narrowtext
\tightenlines

\section{Introduction}
Although the nature of the
superconducting pair wave function in high -$T_c$ cuprates is not yet
known strong
evidences of a major $\rm d_{x^2-y^2}$ symmetry exists 
\cite{Cox,Harlingen,Scalapino}. 
Experiments sensitive to the internal phase structure of the pair
wave function  reported a sign reversal
of the order parameter supporting $d$ wave symmetry \cite{dwave}.
Most recently from various experiments and
theory it appears that 
the pairing symmetry of these family could be a mixed one like
 $\rm d_{x^2-y^2} + e^{i\theta}{\alpha}$ where $\alpha$ could be
something in the $s$ wave family or $d_{xy}$. 
There were early questions from tunneling experiments regarding
 the pure d-wave symmetry\cite{Dynes} 
as the data supports 
 an admixture of d and s-wave components  due to orthorhombicity in
YBCO\cite{Walker,Carbotte}.
Possibility of a minor but finite $id_{xy}$  symmetry alongwith
the predominant $d_{x^2-y^2}$ has also been
suggested\cite{SBL} in connection with magnetic defects
or small fractions of  a flux quantum $\Phi_0=hc/2e$ in YBCO powders. Similar
proposals came from various other authors in the context of magnetic field,
magnetic impurity, interface effect etc. \cite{Laughlin,Krishana,others}
The experimental result by Krishana {\it et al.}, was interpreted 
as a signature of induction of a minor component {\it eg}, $id_{xy}$ or $is$ 
in a $d$-wave superconducture with the application of magnetic field along the
$c$ axis. 

In this work, we study in details the effect of a weak Zeeman magnetic field on
superconductors with mixed order parameter symmetry like
$\Delta (k) = \Delta_{d_{x^2-y^2}} + e^{i\theta}\alpha$ with $\alpha=d_{xy}$,
$s$ for arbitrary $\theta$. It is well known that such superconductors with
$\theta \neq 0$ and $\alpha \neq 0$ 
corresponds to broken time reversal states (BTRS). These
BTRS states lift the
directional degeneracy of charge currents by
admixing a subdominant $\alpha$-wave component to the d-wave pairing 
state and a
spontaneous finite current appears\cite{trsbreview}.
An application of Zeeman magnetic field can lift the spin degeneracy leading
to suppression of BTRS ; a pure $d$-wave occurs with increasing magnetic field.
For mixed symmetry as above with $\theta =0$ that
 preserves time reversal symmetry and are nodeful respond differently to
the Zeeman field as compared to $\theta \neq 0$ states. For nodefull 
$\theta =0$ state, the local gap $\Delta (k)$ of small magnitude 
over the Fermi surface may be destructed with the application of Zeeman field
leading to a paramagnet pocket. This although true for $\theta \neq 0$ states,
but such states correspond to {\em fully gapped} situation all over the
Fermi surface which 
causes {\em weak} response to the magnetic field. A clear picture on the above
will be demonstrated in this article. 
It may be mentioned that the high temperature superconductors are quasi
two dimensional in nature and therefore, a magnetic field parallel to the 
$\rm Cu-O$ plane does not couple to the orbital motion of the electrons in the
plane. Therefore, we shall not consider spin-orbit interaction in this work.

In connection with the discussion of order parameter symmetry in cuprates,
we would further like to mention that the proposal of mixed order parameter
symmetry got the correct momentum when experimental data 
on longitudinal 
thermal conductivity by Krishana {\it et al,} \cite{Krishana}
 of $\rm Bi_2Sr_2CaCu_2O_8$ compounds and 
that by Movshovich {\it et al,} \cite{Krishana} showed supportive indication
to such proposals. There are experimental results related to interface effects
as well as in the bulk that indicates mixed pairing symmetry (with dominant
$d$-wave) \cite{others}, thus providing a strong threat to the pure 
$d$ wave models. There were early orginal works as regards to the modifaction 
of superconductivity due to application of Zeeman magnetic \cite{various} and
very recently, the Zeeman suppression was discussed in mesoscopic systems 
\cite{braun}.

\section{Model Calculation} 

The free energy of a
two dimensional planar 
superconductor with arbritary pairing symmetry in presence of
a magnetic field may be written as,
\begin{equation}
F_{k,k^\prime}(h) = -\frac{1}{\beta} \sum_{k,\sigma = \pm} 
\ln (1 + e^{-\sigma \beta E_{k}^\sigma}) +
\frac{\mid \Delta_k \mid^2}{V_{k k^\prime}}
\label{free}
\end{equation}
where $E_{k}^{\sigma} =  \sqrt{(\epsilon_{k}-\mu)^2 + \mid \Delta_k \mid^2}
+ \sigma (g\mu_B/2)B$
are the energy
eigen values of a Hamiltonian that describes superconductivity,
$(g\mu_B/2)$ is the magnetic moment of the electrons. This includes the 
assumption that the Zeeman field raises/lowers the energy of the spin up/down
quasiparticle states.
We minimize the free energy, Eq.\ (\ref{free}) {\it i.e}, 
$\partial F/\partial \mid \Delta\mid$ = 0, to get the gap equation as,
\begin{equation}
\Delta_k = \sum_{k^\prime} \frac{V_{kk^\prime}}{2} \frac{\Delta_{k^\prime}}
{2 E_{k^\prime}}\left(
\tanh (\frac{\beta E_{k^\prime}^+}{2})+ \tanh (\frac{\beta 
E_{k^\prime}^-}{2})\right)
\label{gapeq}
\end{equation}
where $\epsilon_k$ is the dispersion relation taken from the ARPES data
\cite{Ding} and $\mu$ the chemical potential will control band filling 
through a number conserving equation given below.

Since the applied Zeeman field modifies the SC quasiparticles of spin up and
down differently, their occupation probablities are also modified.
 The number conserving equation that controls the band filling through 
chemical
potential, $\mu$ in presence of Zeeman field is given by, 
\begin{eqnarray}
&&\rho(\mu,T,h)= \nonumber \\
&&
\sum_k\left[1-\frac{1}{2}\frac{(\epsilon_k - \mu)}{E_k}\left(
\tanh\frac{\beta E_{k}^+}{2} + \tanh\frac{\beta E_{k}^-}{2}\right) \right]
\label{dens}
\end{eqnarray}
where $h=(g\mu_B/2)B$.
Let us consider that the overlap of orbitals in different unit cells
is small compared to the diagonal overlap. Then in the spirit of tight binding
lattice description, the matrix element of the pair potential used in
the SC gap equation Eq. (\ref{gapeq}) may be obtained
as,
\begin{eqnarray}
V(\vec q) &=& \sum_{\vec \delta} V_{\vec \delta} e^{i \vec q \vec R_{\delta}}
= V_{0} + V_1 f^d (k) f^d(k^\prime) + V_1 g(k) g(k^\prime) \nonumber \\
&&
 + V_2 f^{d_{xy}}(k)f^{d_{xy}}(k^\prime)
+V_2 f^{s_{xy}}(k)f^{s_{xy}}(k^\prime) 
\label{ppot}
\end{eqnarray}
where in the first result of the equation \ (\ref{ppot}) $\vec R_{\delta}$
locates nearest neighbour and further neighbours, $\vec \delta$ labels and
$V_n$, $n=1,2$ represents strength of attraction between the respective
neighbour interaction. The first term in the above equation $V_{0}$
refers to the on-site interaction which has an effective attractive value
giving rise isotropic $s$ wave. 
The second and third terms are responsible for $d$ and extended $s$ wave
symmetry superconductivity whereas the $4^{th}$ and the $5^{th}$ terms are
responsible for $d_{xy}$ and $s_{xy}$ symmetries respectively. We restrict
only to singlet pairing states ({\it i.e}, $\Delta (k)=
\Delta (-k)$) as applicable for high temperature superconductors.
 The momentum
form factors are obtained as,
\begin{eqnarray}
& &f^d(k) = \cos (k_xa) - \cos (k_ya) \nonumber \\
&&
g(k) = \cos (k_xa) + \cos (k_ya) \nonumber \\
&&
f^{d_{xy}}(k)= 2 \sin (k_xa) \sin (k_ya) \nonumber \\
&&
f^{s_{xy}}(k)= 2 \cos (k_xa) \cos (k_ya)
\label{symm}
\end{eqnarray}
For  two component order parameter symmetries as mentioned above, 
we substitute the required
form of the potential and the corresponding gap structure into the either 
side of Eq. \
(\ref{gapeq}) which gives us an identity equation. 
Then separating the real and imaginary parts together with
comparing the momentum dependences on either side of it we get gap equations 
for the amplitudes in different channels as, 
\begin{eqnarray}
\Delta_j= \sum_k\frac{V_j}{2}\frac{\Delta_jf^{j^2}_{k}}{2E_k}\left[\tanh(
\frac{\beta E_{k}^+}{2})+\tanh(\frac{\beta E_{k}^-}{2}) \right]
\label{gapcomp}
\end{eqnarray}
where $j = 1, 2$ corresponding to two components $d$ and $s$ or $d$ and 
$d_{xy}$ symmetries.
Considering mixed symmetry of the form
 $\Delta (k) = \Delta_{d_{x^2-y^2}}(0)f^d (k) 
+e^{i\theta}\Delta_{s}(0)$ one identifies $\Delta_1 = 
\Delta_{d_{x^2-y^2}}(0)$, $\Delta_2 = \Delta_{s}(0)$,
$f_{k}^1 = f^d (k)$, $f_{k}^2 = 1$ and $V^1=V_1$, $V^2=V_0$ in Eq. (4). 
Similarly, for mixed
symmetries of the form $\Delta (k) = \Delta_{d_{x^2-y^2}}(0)f^d (k)
+e^{i\theta}\Delta_{d_{xy}}(0)f^{d_{xy}}$ 
$\Delta_2 = \Delta_{d_{xy}}(0)$,  $f_{k}^2=f_{k}^{d_{xy}}$ and
$V^1=V_1$, $V^2=V_2$ of Eq.(4). 
The potential required to get such pairing symmetries are discussed
in Eq. (\ref{symm}).

 We solve self-consistently the above three equations (Eq.\ref{gapcomp}
and Eq.\ref{dens}) in order to study the
phase diagram of a mixed order parameter superconducting phase in presence
of Zeeman magnetic field. 
The numerical results obtained for the gap amplitudes through Eqs. 
\ (\ref{gapcomp},\ref{dens}) will be compared with free energy minimizations 
via Eq. \ (\ref{free}) to get the phase diagrams.

\section{Results and Discussions}
 We present in this section our numerical results 
for a set of fixed parameters, {\it e.g},
a cut-off energy $\Omega_c$= 500 K around the Fermi level above which 
superconducting condensate does not exist, a fixed transition temperature
of the minor component $T_{c}^\alpha (h=0) = 24$ K and the bulk $T_c = 85$ K
determined by the $d$-wave order parameter.
In figures 1 and 2 we present results for 
$\Delta (k) = \Delta_{d_{x^2-y^2}}(0)f^d (k)
+ e^{i\theta}\alpha$ symmetries for $\theta = \pi/2$
and $\theta =0$ respectively. Such symmetries would arise from a 
combination of two component pair potentials ($2^{nd},~1^{st}$),
($2^{nd},~4^{th}$) terms in Eq. \ref{ppot}  for $\alpha= s ~\& ~d_{xy}$ 
respectively. The amplitudes of extended $s$-wave states like $s_{x^2+y^2}$,
$s_{xy}$ are found to be finite only towards very low band filling, 
$\rho \sim 0$ for $\theta=\pi/2$ hence does not cause any mixing with 
the predominant $d$-wave. 
Therefore, we shall discuss only the results of 
$\theta =0$ and $\theta =\pi/2$ for $\alpha= s ~\& ~d_{xy}$. These
two phases of $\theta$ can cause important differences. It
is known that for any $\theta \neq 0$, time reversal symmetry is locally
broken \cite{hng2} at lower temperatures
with the onset of the secondary component
 which correspond to a phase transition to an 
fully gapped 
phase for $h=0$, 
from a partially ungapped phase of $d_{x^2-y^2}$ symmetry. On the 
other hand, the $\theta=0$ phase still remains nodeful, although the nodal 
lines shifts a lot from the usual $k_x=k_y$ lines of the $d_{x^2-y^2}$.
In Fig. 1(a) and (b)  we present the temperature dependencies of the order
parameters in the complex mixed symmetry. Curves corresponding to the $d$-wave
channels are represented with thinner joining lines of different styles whereas
 the minor $s$ component is denoted through that of same style but with
thicker lines (same strategy will be carried out in other figures as well). For a zero Zeeman field ($h=0$),
the amplitude of the $d$-wave is suppressed with the onset of the minor 
$s$ component at $T=24$ K which leads to a kink like structure (cf. the thin
solid curve in Fig1 (a)).
\begin{figure}
\epsfxsize=4.5truein
\epsfysize=4.5truein
{\epsffile{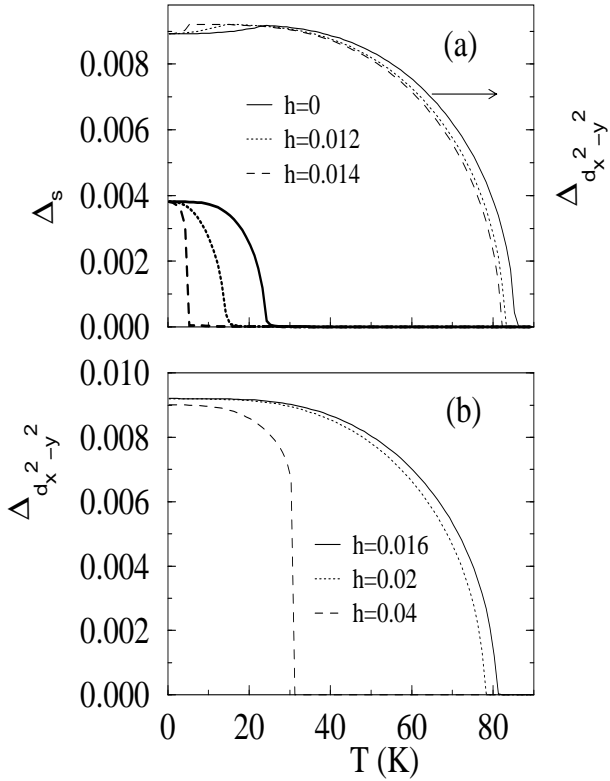}}
\caption{{\bf (a)} 
Amplitudes of the $\rm \Delta_{d_{x^2-y^2}}$ and 
$\rm \Delta_{s}$ at a
fixed band filling $\rho = 0.85$ as a function of temperature
(in K) for $\rm \theta = \pi/2$ ({\it i.e}, 
$\rm d_{x^2-y^2}+is$) phase in various values of
Zeeman field ($h$). With the application of weak magnetic field
while the $\rm d_{x^2-y^2}$ remains almost unaffected
the transition temperature ($T_{c}^\alpha$) 
of the minor $\alpha =s$ component is suppressed strongly. At a field
value of $h = 0.014$ eV the $s$-wave shows a first order transition. 
{\bf (b)} For field values $h \geq 0.016$ eV the minor $s$-channel is
completely suppressed leading to a pure $d$ wave order parameter. The
thermal behaviour of $d$-wave superconductivity in presence of the
Zeeman field is presented in Fig.1(b). While the amplitudes at 
$T=0$ K remains almost unaffected with field, the transition temperatures get
affected.  A magnetic field induced first 
order transition is observed at $h = 0.04$ eV.}
\label{fig:dsstar-complex}
\end{figure}

 With application of weak Zeeman field the
transition temperature of the minor $s$ state decreases while the zero 
temperature magnitude remains the same. This causes a shift in the kink like
structure in the thermal dependence of the $d$ wave channel towards
lower temperature with increasing
field and hence a small enhancement in the $d$-wave with field at 
lower temperature occurs. This point will
be clearer from Fig. 5(b) as discussed latter. At a field value 
$h_c = 0.016$ eV, the $s$ wave component is completely suppressed leading to a 
pure $d$ wave phase. Thus we have a magnetic field induced transition at lower
temperature from {\em fully gapped phase} of the $d+is$ state to a 
{\em partially} gapped phase of the $d$-wave. Therefore, in absence or very
low magnetic field, there is a transition from
 a {\em partially gapped} phase of the $d$-wave at higher temperature
to a {\em fully gapped} $d+is$ phase at lower temperatures. 
With increasing field the fully gapped 
phase region with respect to temperature decreases and brings back the ungapped
phase of the $d$-wave. These phase transitions will have important bearings
in the thermodynamic and transport properties. 
\begin{figure}
\epsfxsize=3.25truein
\epsfysize=3.25truein
{\epsffile{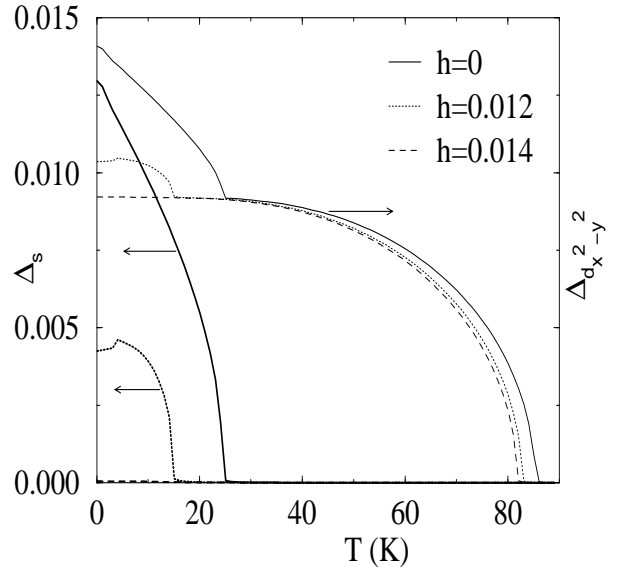}}
\caption{Same as that of figure 1(a) except $\rm \theta = 0$ ({\it i.e,}
$\rm d_{x^2-y^2} + s$ symmetry). For the $d$-wave channel, its amplitude
is found to increase below $T_{c}^s$ in contrast to $\theta = \pi/2$ case
where the $d$-wave amplitude is suppressed. The $s$-wwave channel suffers
 drastic suprresion in $T_{c}^s$ as well as the zero temperature amplitude
in contrast to that in Fig. 1(a). Note, the critical field ($h_c$) at which 
the $s$ component is completely suppressed is 0.013 eV in contrast to 
0.016 eV in case of $\theta = \pi/2$ (cf. Fig1a).}
\label{fig:dsstar-real}
\end{figure}
In Fig. 1(b) we present the paramagnetic state of the $d$-wave. This phase 
has also been qualitatively investigated recently by Yang and Sondhi 
\cite{yang} with possibility of pairing with finite momentum. 
We therefore restrict to present only
the details thermal dependence of the $d$-wave superconductivity in presence
of Zeeman magnetic field that were not discussed. 
We show that with increasing field (at lower fields)
the zero temperature magnitude of the $d$-wave remains unchanged whereas the 
$T_c$ is reduced. At higher field {\it e.g}, $h=0.04$ one sees a first order 
transition from superconducting state to the normal state with respect to
temperature. This behaviour causes a magnetic field induced enhancement of
the $2 \Delta/k_BT_c$ ratio. This ratio is crucial for many physical properties
like specific heat jump etc. and hence expected to have drastic effect with 
magnetic field. 
\begin{figure}
\center
\epsfxsize=3.25truein
\epsfysize=3.25truein
\leavevmode
{\epsffile{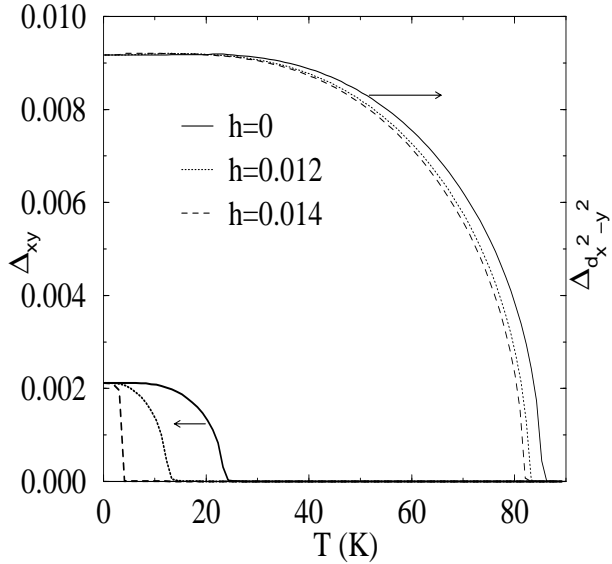}}
\caption{Amplitudes of the $\rm \Delta_{d_{x^2-y^2}}$ and
$\rm \Delta_{d_{xy}}$ at a
fixed band filling $\rho = 0.85$ as a function of temperature
(in K) for $\rm \theta = \pi/2$ ({\it i.e},
$\rm d_{x^2-y^2}+id_{xy}$ in various values of
Zeeman field ($h$). With the application of weak magnetic field
while the $\rm d_{x^2-y^2}$ remains practically unaffected
the transition temperature ($T_{c}^\alpha$)
of the minor $\alpha =d_{xy}$ component is suppressed strongly.
Similar to that of Fig. 1a, a first order transition is seen in the
minor $d_{xy}$ channel for $h=0.014$ eV.}
\label{fig:dsxy-complex}
\end{figure}
In Fig.2 we describe the thermal behaviour of superconductors with
$d+s$ symmetry in presence of magnetic field. First of all, in absence
of magnetic field, the thermal behaviors at lower temperatures is quite
different (as we have seen in Fig. 1), 
the $s$-wave gap opens very fast below $T=24$ K and also
induces a growth to the $d$ at a faster rate than that above $T=24$ K.
While the $s$-wave has about three times the zero temperature value compared
to $d+is$ phase, the $d$ wave also have quite larger value. With application
of a very small magnetic field the $s$-component is suppressed largely; both
its $T_c$ and the zero temperature gap being suppressed. The $d$-wave gap
magnitude is also suppressed while its $T_c$ remained practically unchanged
with such small values of the magnetic field. More importantly, although
both the $d$ and the $s$ channel has larger magnitude in the $d+s$ phase the
critical field ($h_c =0.013$ eV) 
at which the $s$ component vanishes completely is {\em smaller} compared
to that for the $\theta = \pi/2$ phase ($h_c =0.016$ eV) of the mixed symmetry. 
 Distinctly, the response of the Zeeman field to the $d+s$ superconductors
is more pronounced than the $d+is$ superconductors. This study therefore also
revealed the importance of the phase of the minor component in $d$-wave
superconductors with a pratical
example of effect of Zeeman 
magnetic field, for the first time. In order to establish
the importance of the phase of the minor component we also study the
effect of Zeeman magnetic field in $\rm d_{x^2-y^2}+id_{xy}$
and $\rm d_{x^2-y^2}+d_{xy}$ symmetry superconductors respectively. 
The qualitative as well as quantitative behaviors remain almost same as that  
shown in figures 1 and 2. Thus establishing that the response of the
superconducors with mixed order parameter symmetries with $\theta=0$ phase of
the minor component is stronger.
\begin{figure}
\epsfxsize=3.25truein
\epsfysize=3.25truein
{\epsffile{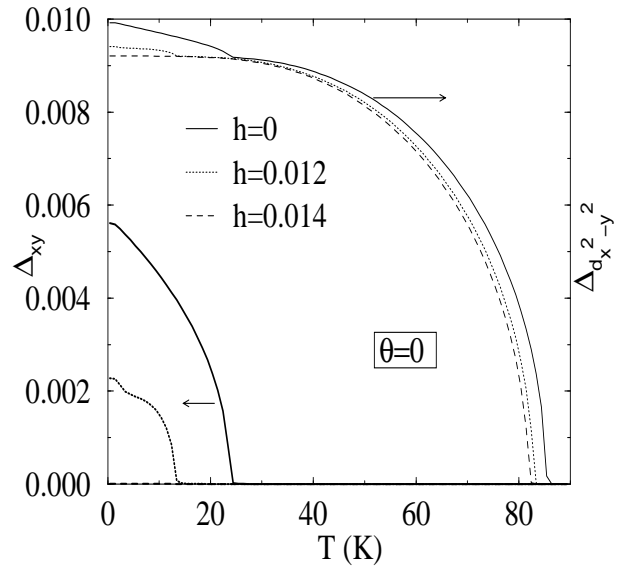}}
\caption{ Same as that in figure 3 except for $\rm \theta=0$ {\it i.e}
$\rm d_{x^2-y^2}+d_{xy}$ phase that preserves the time reversal symmetry.
The notable difference is that the minor $\rm d_{xy}$ 
component although have a magnitude at $T=0$ K and $h = 0$
three times larger than that for $\theta = \pi/2$ case, the minor component
is suppressed at a lower critical field $h_c = 0.013$ eV.}
\label{fig:dsxy-real}
\end{figure}
In figures 5 and 6 we concentrate on studying behavior of such 
mixed order parameter symmetry superconductors in presence of 
magnetic field as a function of band filling ($\rho$) for $d_{xy}$ and $s$ as
minor components respectively. With application of a small field the 
minor component $\alpha=d_{xy}$ or $s$ is suppressed
 only at lower fillings and then with increasing field it is suppressed
in the optimal doping regime suddenly. 
Therefore, with increasing field one finds
mixed symmetry region as well as pure $d$-wave region with respect to filling
\--- a {\em non-uniform} superconductivity.
This behavior depends on the nature of the minor component. For $\alpha =s$,
the mixed symmetry is possible around half-filling and at around $\rho=0.7$
in an intermediate field value.
With increasing field mixed symmetry regions around both the band fillings 
shrinks and at a field value of $h=0.03$ the $s$-wave vanishes leading to a
pure $d$-wave phase. It may be noticed the $d$-wave channel is enhanced (cf.
Fig 5(b)) in
the optimal doping region 
with intermediate field values as was also mentioned 
earlier while discussing Fig1 (a). This however does not occur
in case of $d_{xy}$. For $d_{xy}$ minor component, 
the minor component is suppressed strongly 
only from the lower filling with
increasing field and at an intermidiate field the mixing is possible only
near half filling. The $d$-wave boundary towards lower fillings as well
as around half-filling also shrinks
with increasing field values in contrast to $\alpha=s$. One always see a sharp
transition from a mixed phase to a pure $d$-wave phase
 or otherway around with respect to
band filling irrespective of the minor component in a given magnetic field. 
 At larger fields
 ($h > 0.03$) when the minor component is suppressed
completely {\it i.e}, one has a pure $d$-wave phase, the $d$-wave phase
also show sharp transition from $d$-wave superconducting state to a normal
state (cf. Fig. 6).
\begin{figure}
\epsfxsize=4.5truein
\epsfysize=4.5truein
\leavevmode
{\epsffile{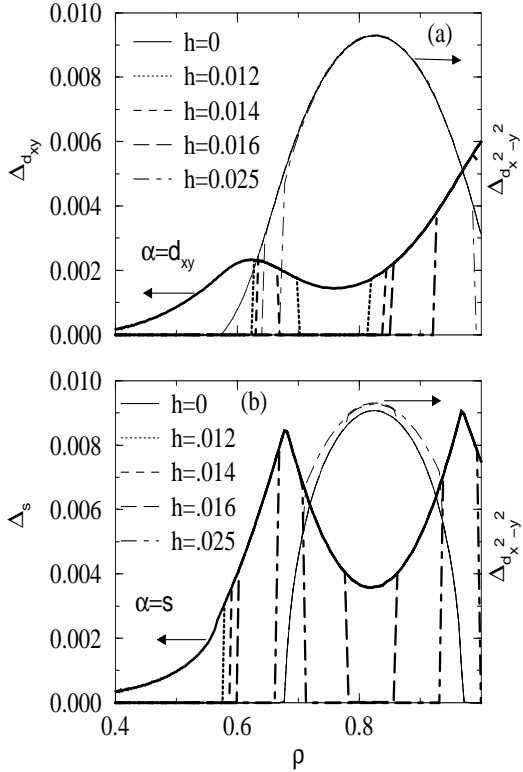}}
\caption{Zero temperature amplitudes of the different pairing channels as
a function of band fillings in the $\rm d_{x^2-y^2} + e^{i\theta}{\alpha}$
picture for several fields with $\theta = \pi/2$. An increase in the amplitude
of the $d$-wave channel with some particular field values in the 
$\rm d_{x^2-y^2} + is$ picture near optimal doping is worth noticing. 
With field increasing the order parameter symmetry changes to pure $d$-wave at
optimal and larger doping.
}
\label{fig:dsxy.temp}
\end{figure}
In Fig. 6 we study the same as that in Fig. 5 for $d+s$ and $d+d_{xy}$ 
symmetries. In contrast to the $\theta =\pi/2$ phase of the minor component
(as in Fig. 5), the minor components have very large values in the $\theta =0$
phase although have the same $T_{c}^\alpha$ as mentioned earlier, in absence
of the magnetic field. With very small magnetic field such large amplitudes of
the condensation in the minor channel is strongly suppressed (see Fig 6(b) 
specially). The nature of
suppression, as in Fig.5, is different for different minor component. For 
example, $d_{xy}$ is suppressed only from the lower filling whereas the $s$-wave
is suppressed both from the lower filling as well as around the optimal doping.
\begin{figure}
\center
\epsfxsize=4.5truein
\epsfysize=4.5truein
\leavevmode
{\epsffile{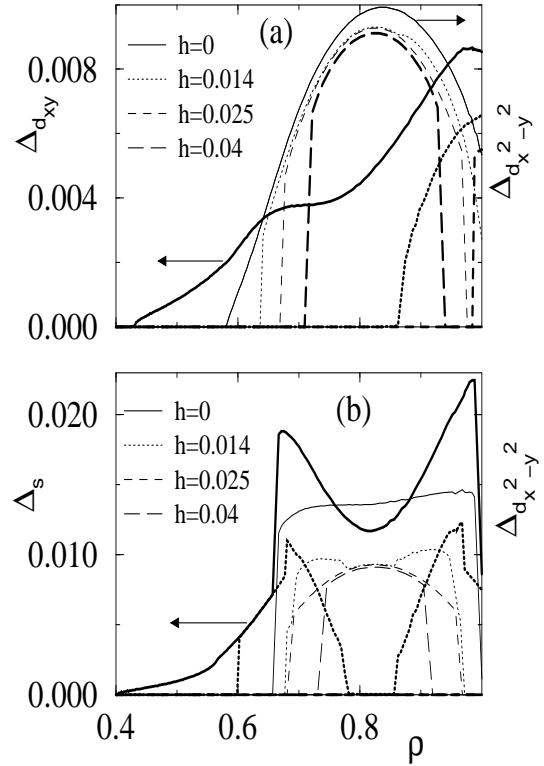}}
\caption{Same as figure 5 except $\theta = 0$. Stronger and
rapid suppression of the minor channel with field in contrast to that
in Fig. 3. A visible change in the dominant $d$-wave channel is also seen
in contrast to that in Fig. 3. Note, in case of $d+s$ picture, {\bf (b)},
the $d$-wave channel shows plateau with respect to band filling.}
\label{fig:dsstar.temp}
\end{figure}

Overall, it is very distinct that the magentic field affects the $\theta =0$ 
mixed phase more stongly than the $\theta =\pi/2$ phase. This is due to
the fact that the response of applied Zeeman field  is paramagnetic with
destruction of superconductivity over parts of the Fermi surface where
the Zeeman field exceeds the local magnitude of the
$k$-dependent gap resulting a spin
polarization in the normal electrons. For $\theta = \pi/2$ phase, the nodes
are missing and the Fermi surface is gapped all over, although the gap will
have local minima. Thus response of the $\theta =\pi/2$ phase is weaker
compared to the $\theta = 0$ phase. Noticiably, the critical value at which
the minor component vanishes completely in all band fillings
 for the $\theta =0$ phase is $h_c = 0.025$ eV whereas that for 
$\theta =\pi/2$ is  $h_c = 0.03$ eV. (The value of $h_c$ depends on $\rho$,
as well as $\alpha$ and $\theta$).

Above $h_c=0.025$, the paramegnetic state of the $d$-wave superconductors
are also presented. The region of $d$-wave superconductivity shrinks with
increasing field around the band filling $\rho = 0.82$. The change from
$d$-wave to spin polarized normal state is very sharp with respect to the
band filling. In the $d+s$ state a plateau has been observed in the $d$ channel
in weak or zero field, similar to the behaviour known for the YBCO 
systems \cite{19}.
It may be mentioned that Krishana {\it et al.}, found
 the longitudinal thermal
conductivity of $\rm Bi_2Sr_2CaCu_2O_{8+\delta}$  at lower temperatures (5K to
20 K) decreases with the increase in magnetic field applied along $\rm c$-axis.
 Above a
critical value of the magnetic field $\rm H_k(T)$, the
 thermal conductivity
cease to change with the magnetic field and develops a plateau.
It was proposed
\cite{Krishana} that the $\rm d_{x^2-y^2}$ pairing state is unstable against
the formation of $\rm d + e^{i \theta}\alpha$ (where $\rm \alpha = s, d_{xy}$)
in presence of $\rm H_k (T)$ such that the loss of
quasiparticle transport in the thermal conductivity can be explained.
In contrast, we started with a $\rm d + e^{i \theta}\alpha$ picture and
application of Zeeman field (which may be mapped as application of magnetic
field parallel to the 2D $\rm Cu-O$ plane) without considering the orbital
effect did not find any enhancement in the condensation of the minor channel but
suppression leading to a pure paramegnetic $d$ state.  
\section{Summary}
In summary we have performed a detailed study on the effect of Zeeman
magnetic field on mixed pairing symmetry with predominant $d$-wave which
seems to be very promising symmetry for the high $T_c$ systems. Thus we have
described the paramagnetic state in the mixed symmetry superconductors and
subsequently in the $d$-wave superconductors.
In particular we established that the phase of the minor component mixed
with predominant $d$-wave is of immense importance. The $\theta =0$ phase 
minor component symmetry responds to Zeeman field more profoundly than 
the $\theta =\pi/2$ of the minor component. It will be very interesting to
calculte the specific heat, Magnetization, density of states as a function of
magnetic field using this model. We argued that the orbital effect is 
secondary when a magnetic field is applied parallel to the conducting plane.
This may indicate that the experimental observation by Krishana {\it et al.},
involve strong coupling of spins to orbitals due to application of magnetic
field perpendicular to the plane at lower temperatures.    
However, the order parameter alone does not completely determine the
thermodynamic property of a system. In Ref. \cite{yang}, 
some estimations and scaling relations in the 
change of various physical properties due to Zeeman field are given 
for a
pure $d$-wave superconductor. It turns out, a weak
Zeeman field does little to the order parameter but may
profoundly affect the thermodynamic property of a pure $d_{x^2-y^2}$
supercondcutor. Such effects will presumably also remain for mixed
superconductors with $\theta = 0$ as the gapnodes similar to $d_{x^2-y^2}$
 remains, its effect
in the $\theta =\pi/2$ phase has to be studied more carefully.
\section{Acknowledgments}
A large part of this work was carried out at the Instituto de F\'isica, 
Universidade Federal Fluminense, Brazil and was
financially supported by the Brazilian funding agency FAPERJ, project no. 
E-26/150.925/96-BOLSA.

\end{multicols}
\end{document}